\documentclass[aps,prl,reprint,superscriptaddress,nofootinbib,nobibnotes,floatfix]{revtex4-2}
\usepackage{graphicx}
\usepackage{bm}
\usepackage{amssymb}
\usepackage{amsbsy}
\usepackage{amsmath}
\usepackage{mathtools}
\usepackage{xcolor}
\usepackage{stmaryrd}
\usepackage{mathrsfs}
\usepackage[mathscr]{euscript}  
\usepackage{tikz}
\usetikzlibrary{shapes}
\usepackage{cancel}
\usepackage{booktabs}
\usepackage[normalem]{ulem}
\usepackage{etoolbox}
\usepackage{siunitx}
\usepackage{titlesec}
\usepackage{titlecaps}

\titleformat{\section}{\normalfont\bfseries}{\thesection}{1em}{}
\titleformat{\subsection}{\normalfont\bfseries}{\thesubsection}{1em}{}
\titleformat{\section}{\normalfont\bfseries\filcenter}{\thesection}{1em}{}
\titleformat{\subsection}{\normalfont\bfseries\filcenter}{\thesubsection}{1em}{}
\titleformat{\subsubsection}{\normalfont\bfseries\filcenter}{\thesubsubsection}{1em}{}



\renewcommand{\figurename}{Fig.}

\begin{document}

\title{Enzyme active bath affects protein condensation}
\author{Kevin Ching}
\author{Anthony Estrada}
\author{Nicholas M Rubayiza}
\author{Ligesh Theeyancheri}
\affiliation{Department of Physics, Syracuse University, Syracuse, NY, USA} 
\author{Jennifer M. Schwarz}
\affiliation{Department of Physics, Syracuse University, Syracuse, NY, USA} 
\affiliation{Indian Creek Farm, Ithaca, NY 14850 USA}
\author{Jennifer L Ross}
\email[To whom correspondence should be addressed. E-mail:]{jlross@syr.edu}
\affiliation{Department of Physics, Syracuse University, Syracuse, NY, USA} 

\begin{abstract}
\noindent We investigate how an active bath of enzymes influences the liquid–liquid phase separation (LLPS) of a non-interacting condensing protein. The enzyme we choose to use as the active driver is urease, an enzyme that has been shown by several groups to exhibit enhanced diffusion in the presence of its substrate. The non-interacting LLPS protein is ubiquilin-2, a protein that condenses with increasing temperature and salt. Using a microfluidic device with semipermeable membranes, we create a chemostatic environment to maintain the substrate content to feed the enzymatic bath and remove the products of the chemical reaction. Thus, we isolate the physical enhanced fluctuations from the chemical changes of the enzyme activity. We also compare the results to controls without activity or in the presence of the products of the reaction. We find that the active bath is able to enhance droplet size, density, and concentration, implying that more ubiquilin-2 is in condensed form. This result is consistent with an interpretation that the active bath acts as an effective temperature. Simulations provide an underlying interpretation for our experimental results. Together, these findings provide the first demonstration that physical enzymatic activity can act as an effective temperature to modify LLPS behavior, with implications for intracellular organization in the enzymatically active cellular environment.

\vspace{0.5\baselineskip}
\centering{Active bath \textbar\ Non-equilibrium statistical mechanics \textbar\ Active matter \textbar\ Nanoscale}
\end{abstract}

\maketitle

\begingroup
\renewcommand\thefootnote{}
\setlength{\footnotesep}{0.8\baselineskip}
\footnotetext{Author Contribution: KC performed experiments, analyzed data, and drafted figures and manuscripts. AE and NR assisted in experimental design, performed experiments, and analyzed data. LT performed simulations on model systems and drafted the manuscript. CC helped with experimental design and provided expertise on protein systems, JS advised on simulation and model design, edited the manuscript, and JLR conceived of the experiments, advised on experimental design and analysis, edited figures, and the manuscript.}

\footnotetext{Authors have no competing interests.}
\endgroup

\noindent Biological systems are active and self-regulating, composed of a myriad of interacting components including the cytoskeleton, nuclear envelope and pores, DNA, and organelles, with or without membranes, both outside and inside the nucleus. These energetic, multifarious interactions are dynamic and controlled through enzymes that can alter the local chemistry affecting the interaction between proteins in that space, as well as the mobility of the objects, creating an active bath \cite{parry2014bacterial,guo2014probing}. \\

\noindent Here, we discover that enzymes are actually multi-mode tuning knobs with the capacity to affect the interaction between proteins not just chemically but via their inherent activity, or the nontrivial motion of the enzymes given their ability to convert chemical energy into work. To be specific, it has been shown that enzymes, in the presence of their substrate, can collectively propel micronscale particles \cite{Valles2022,Hortelao2021} and create flows \cite{Sengupta2014,Ortiz-Rivera2016}, and thus be harnessed for propulsion. If many enzymes can be gathered to propel a large colloid, it stands to reason that a single enzyme is also a propulsive unit. Unlike active colloids with many enzymes coupled together, an enzyme is a single propulsive element and thus will not offer as much thrust. Further, the small size allows for rapid reorientation of the enzyme compared to the large colloid. Thus, the Pecl\`et number, $Pe$, of an enzyme will be far less than one, causing the enzyme to appear to diffuse faster instead of propel. Indeed, many fast, ergonic, exothermic enzymes have been shown to display such ``enhanced diffusion'' \cite{Zhao2018,Börsch1998,Jee2020}, as would be expected, though the exact physical mechanism remains elusive. To this end, enzymes have been proposed to serve as nanoscale active matter implying that the insides of cells are active baths \cite{guo2014probing,parry2014bacterial,Demarchi2023,ghosh2021}. \\

\noindent Directly isolating the physical effects of enzymatic activity on whole-cell organization is daunting, so we focus on reconstituting protein organization via liquid–liquid phase separation (LLPS), which underlies membraneless organelles \cite{Feric2016,Brangwynne2009,Vanderweyde2013,kedersha2000}. To do so, we create a well-controlled active bath of enzymes consisting of urease, which has been shown to perform enhanced diffusion \cite{Muddana2010,Xu2018,Riedel2015}. The active bath is chemostatically isolated using an experimental chamber that can both feed the enzymes to maintain the activity and remove the products of the reaction \cite{Park2016}.  As for the LLPS protein, we study a truncated version of ubiquilin-2 that has been shown to condense into droplets with increasing temperature and ionic strength \cite{Yang2019}. We use the phase diagram of the ubiquilin-2 protein as a read-out of the effective temperature imparted to the system by the enzyme activity. We find that the activity of enzymes enhances the phase separation of proteins into the condensed state increasing droplet density, number, and protein concentration within the droplets. Coarse-grained simulations of sticker–spacer polymers in an active bath, whose particles are approximately four times the size of the polymer chains, exhibit enhanced cluster growth, increased polymer partitioning, and polymer expansion within droplets, supporting the experimental results. As the activity increases in the simulations, the active bath particles help corral the polymers so that the polymers interact more effectively.  \\

\noindent 
Together, these results show that an enzymatically driven active bath can physically remodel the condensed state of proteins, effectively tuning organization and material properties, thereby expanding the tunability landscape beyond purely chemical control. This points toward strategies for bio-inspired active materials that assemble hierarchically from the nanoscale to the macroscale.

\section*{Results}
\begin{figure}[h!tbp]
\centering
\includegraphics[width=0.5\textwidth]{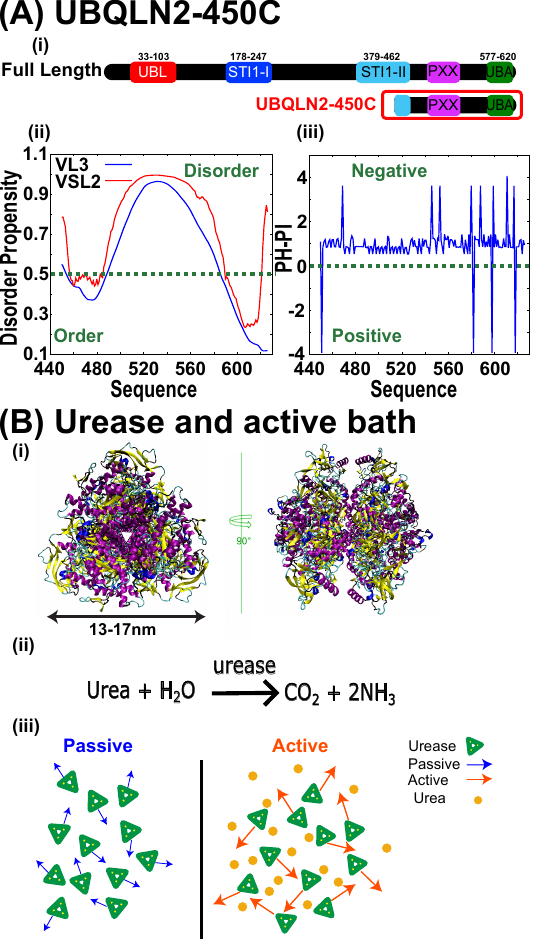}
\caption{\textbf{Model systems for protein phase separation and enzyme activity.} (A) Ubiquilin-2 is a model phase separating protein. (i) Schematic of full length ubiquilin-2 (top), and the truncated form from 450-624 aa used in this study (bottom). (ii) Disordered propensity of UBQLN2-450C. (iii) Charge of UBQLN2-450C, at pH 6.8. (B) Urease is the enzyme used to create an active bath. (i) Crystal structure of the urease hexamer (PDB \# 3LA4 visualized with VMD \cite{HUMP96}). (ii) Chemical reaction of urea hydrolysis catalyzed by urease \cite{Kot2003}. (iii) Schematic of active bath driven by urease enhanced diffusion. Without urea, the system is thermally-driven with no activity (left). With urea, the enzyme activity enhances the diffusion of urease to create an active bath (right).}
\label{fig:system}
\end{figure}

\subsection*{Chemostatic control to isolate physical nature of active bath} 
\noindent We seek to investigate if an ``active bath'' of enzymes can control the phase of a well-characterized protein that undergoes liquid-liquid phase separation. Our model phase-separating protein is a truncated form of ubiquilin-2, specifically the c-terminal end from amino acid 450 to 624 (UBQN2-450C, Fig. \ref{fig:system}Ai), which is partially intrinsically disordered (Fig. \ref{fig:system}Aii) and is mostly negative at neutral pH (Fig. \ref{fig:system}Aiii). The truncated form has been previously shown to display a lower critical solution temperature (LCST), condensing into droplets above $30^{\circ}C$. The protein condenses with increasing salt concentration, likely from screening the negative charges and altering the hydrophobic interactions \cite{Yang2019,Dao2018,Dao2019}. \\

\noindent Our model enzyme to create the active bath is urease, which catalyzes the breakdown of urea into carbon dioxide and ammonia (Fig. \ref{fig:system}B). This hexameric protein is exogonic, exothermic, and has a fast turn-over rate at saturating reactant conditions \cite{krajewska2012,Das2002}. Several groups have shown that it displays enhanced diffusion, a process where the enzyme chemical reaction appears to increase the diffusion coefficient of the molecule \cite{Riedel2015,Xu2018,Zhang2019,Jan-Philipp2018}. Although the mechanism controlling enhanced diffusion is unknown, the concept that the enzymes may act as nanoscale active matter particles makes this system a particularly interesting one to explore the concept of an active bath and its effects on other biological activities, such as LLPS. \\

\begin{figure*}[h!tbp]
\centering
\includegraphics[width=0.96\textwidth]{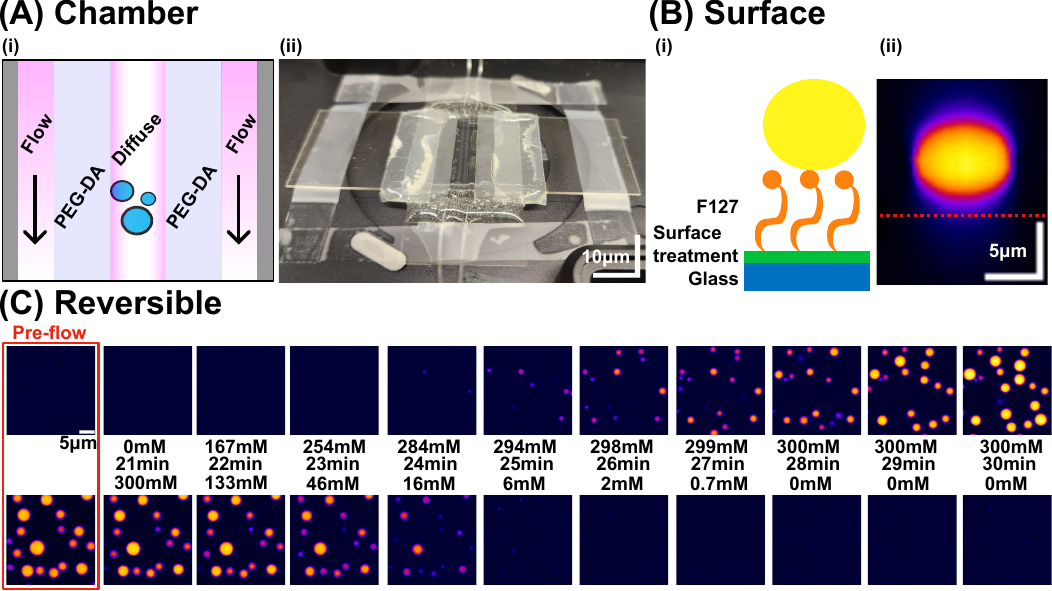}
\caption{\textbf{Microfluidic device for chemostatic control.} (A) Microfluidic chamber with semipermeable membrane walls. (i) Schematic of chamber with three lanes: two outer lanes allow flow to deliver small molecules and remove enzyme catalysis products. The center lane has the experimental system of proteins. (ii) Photograph of the microfluidic chamber. (B) Surface treatment of the chamber. (i) Triblock co-polymer, Pluronic F127 is used to block the surface to inhibit adhesion of UBQN2-450C. (ii) X-Z image of a UBQN2-450C droplet shows no adhesion to the surface. The red, dashed line denotes the surface of the chamber. (C) Microfluidic control of UBQN2-450C phase transition driven with NaCl. Time series showing the formation of condensed UBQN2-450C droplets when 300 mM NaCl is flowed in the outer lanes (top). Condensation is reversed when 0 NaCl is flowed in the outer lanes (bottom). Salt concentration is estimated via 1D diffusion equation with D=$1.958 \times 10^{-9}$ $m^2/s$ \cite{Ghaffari2013}.}
\label{fig:chamber}
\end{figure*}

\noindent To test the effect of increased urease mobility, but not the chemical reaction, we must control the chemical environment of the system. We do this using a microfluidic chamber with three lanes separated by semipermeable membrane walls that allow the passage of small molecules (Fig. \ref{fig:chamber}A) \cite{Park2016}. The inner lane does not experience flow and traps the protein samples inside. The outer lanes allow the addition of 1 mM of urea for the urease enzyme while also removing the products of the reaction. The chamber surface is coated with a triblock polymer, Pluronic F127, to inhibit the phase separated proteins from binding to the surface-treated glass used to create the chamber (Fig. \ref{fig:chamber}Bi). The droplets only loosely associate with the glass to inhibit rapid diffusion, but do not adhere to the surface, as visualized in z-scans of the droplet (Fig. \ref{fig:chamber}Bii). Using this chamber, we can drive the reversible phase transition of the UBQN2-450C by changing the salt concentration from 0 mM to 300 mM and back (Fig. \ref{fig:chamber}C). \\

\noindent In addition to controlling the environment, we compare all active bath experiments to negative and positive controls. Specifically, the negative control is one that lacks activity where the urease enzyme is present without urea. The positive chemical control has urease plus the products of the urease reaction, namely 1 mM carbon dioxide and 2 mM ammonia. The reaction products were created by urease activity being run to completion to generate the products in solution. We found that the addition of sodium bicarbonate or other chemicals that mimic the products also have added salts that altered the ionic strength in solution, artifactually changing the protein condensation. For each experiment, we will compare the active bath to the non-active system and the same system with products to ensure that the changes we observe are only due to the physical activity and not chemistry changes. 

\subsection*{The active bath increases average droplet size}
\begin{figure}[h!tbp]
\centering
\includegraphics[width=0.5\textwidth]{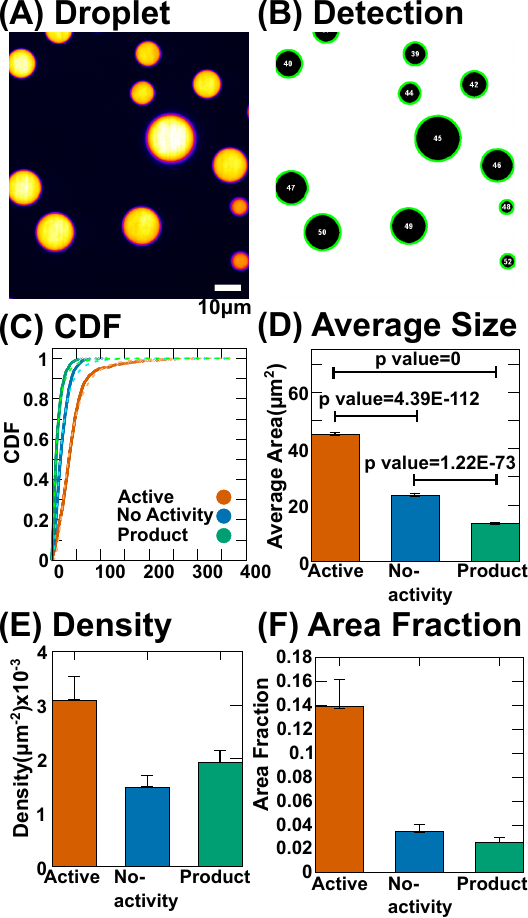}
\caption{\textbf{Droplet size, number density, and area fraction.} (A) Fluorescence image of droplets shows a single slice of the z-scan. (B) Droplet detection using automated particle analysis. (C) CDFs of maximum cross-sectional area of droplets for active bath (orange-filled circles), no activity control (blue-filled circles), and product control (green-filled circles). The CDF is fit with a log-normal CDF function (dash lines). (D) Average droplet area obtained from CDF fit for active condition (orange bar), no activity control (blue bar), and product control (green bar). Error bars represent SEM. Reported p-values calculated using the KS test. (E) Droplet density of active condition (orange bar), no activity control (blue bar) and reaction product control (green bar).  (F) Area fraction of droplets calculated with density and average droplet sizes (Supp. Eqn.~S2) for active condition (orange bar), the no activity control (blue bar), and the product control (green bar). For the density and area fraction, error bars represent the systematic error due to under-counting.}
\label{fig:SDA}
\end{figure}

\noindent Using spinning disc microscopy, we take z-scans of droplets for each experimental condition to quantify the size of the droplets (Fig.~\ref{fig:SDA}A). The maximum cross-sectional area of each droplet is extracted automatically using particle detection (Fig.~\ref{fig:SDA}B, see Supp. Methods for details). The droplets in the active bath are larger compared to both the no activity control and the size of droplets in the presence of reaction products (Fig.~\ref{fig:SDA}C,D). Distributions of the droplet sizes are compared using cumulative distribution functions (CDFs, Fig.~\ref{fig:SDA}C), and each is fit to a log-normal distribution to determine the mean, which we plot as a bar chart (Fig.~\ref{fig:SDA}D, see Supp. Table~S1 for details). We use the Kolmogorov-Smirnov statistical test (KS Test) to compare the distributions and find that all three conditions are statistically distinct from one another (p-values given in Fig.~\ref{fig:SDA}D). The increased size of the droplets in the active bath could imply that the phase diagram is shifted, or it could just imply that the surface tension is altered between these conditions. Indeed, previous studies showed that in a aqueous two-phase system, surface tension drives coalescence and Ostwald ripening \cite{Naz2024,Sarkany2023}. Thus, we need more information to determine if the enzyme activity is acting as an effective temperature to shift the phase diagram or just affecting the droplet material properties. 

\subsection*{The active bath increases droplet density}
\noindent To further quantify the phase of the condensed UBQL2-450C, we measure the number of droplets, in addition to the droplet size. We quantify the droplet number density by counting the number of droplets in z-scans and dividing by the total area observed to normalize against different experiments with variable image numbers (Fig.~\ref{fig:SDA}E). As with the droplet size, we see that the active bath also has higher number density of droplets compared to the control without activity and the control with enzymatic products (Fig.~\ref{fig:SDA}E). There is systematic uncertainty of the measurement because of the finite scanning area, circularity cut-offs, and the fact that droplets on the edge are excluded from the analysis, so the data represents a lower bound for the droplet density (see Methods for details). \\

\noindent When we compare between conditions, we find the droplet density in the active bath is about $2$ times higher than the no-activity control, and $1.4$ times higher than the product control. Unlike the droplet area, the product control has higher density than the no-activity control ($1.3$ times higher, Fig.~\ref{fig:SDA}E, see Supp. Table~S1 for detail). This implies that not only does the active bath increase droplet size, but also the number of droplets formed, suggesting that more UBQLN2-450C is condensing into droplets. \\

\noindent To quantify the amount of UBQN2-450C in the condensed phase, we multiply the size and the number of droplets to calculate an area fraction that the droplets cover. We can also sum up individual droplet size and divide by the total observed area for comparison; both approaches yield similar results (see Supplement for details). The area fraction of the active bath is about $4$ times higher than the no-activity control and $5.4$ times higher than the product control. The no-activity control has slightly higher area fraction, about $1.35$ times higher, than the product control (Fig.~\ref{fig:SDA}F, see Table~S1 for details). These results suggest that the condensed phase represents more of the reaction volume when there is an active bath of urease. For all these experiments, we are assuming that the density of molecules in the condensed phase are identical for all conditions, but that may not be the case. To determine if there is truly more UBQLN2-450C in the condensed phase with the active bath, we must quantify the total amount of condensed protein.  \\

\subsection*{The active bath increases the amount of protein condensed into droplets}
\begin{figure*}[h!tbp]
\centering
\includegraphics[width=1.0\textwidth]{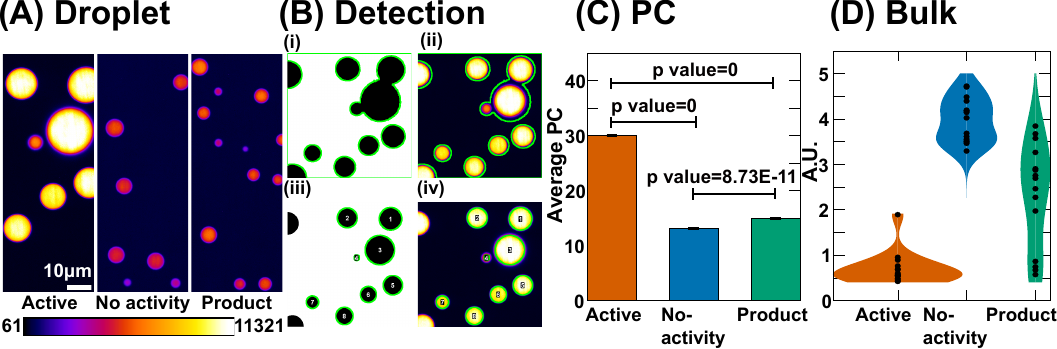}
\caption{\textbf{UBQL2-450C partition coefficient (PC) quantified for all conditions.} (A) Fluorescence image of droplets comparing the active condition (left), the no activity control (middle), and the product control (right) displayed at the same size (scale bar is 10 $\mu m$ or all images) and the same look-up table (AU 61-11321). (B) Quantification method to determine the partition coefficient. (i) Detection and selection of the background region around the droplets to quantify the background intensity with (ii) overlay on the original droplet image. (iii) Detection and selection of the droplets to quantify the droplet intensity with (iv) overlay on the original droplet image. For all images, the green line represents the selection boundary. (C) The average partition coefficient (PC) obtained from CDF fits (Supp. Fig.~S1A). (D) Bulk measurement of the dilute phase using spectrophotometry after centrifugation to remove the condensed phase (N = 15 for each condition). Higher absorbance (A.U.) reports higher UBQLN2-450C concentration.}
\label{fig:PC-640}
\end{figure*}

\noindent To determine if more UBQLN2-450C protein is condensing into the droplets, we need to examine the density of protein within the droplets. Since the UBQLN2-450C is labeled, the intensities of the droplets and background are proportional to the protein concentration of the dense and dilute phases, respectively. Using the z-scans of the droplets with the same exposure time and look-up table, it is already qualitatively obvious that the active case has brighter droplets in addition to larger droplets (Fig.~\ref{fig:PC-640}A). We automatically quantify the intensity of each droplet and the background level (Fig.~\ref{fig:PC-640}B) and measure the ratio of the droplet to background intensities, defined as the partition coefficient (PC). We plot the distribution for each condition and fit with a Gaussian or log-normal function to determine the average partition coefficient (Fig.~\ref{fig:PC-640}C, see Supp. Fig~S1i for CDFs fit). As is qualitatively clear from the images (Fig.~\ref{fig:PC-640}A), the partition coefficient shows that there is significantly more UBQN2-450C in droplets with activity compared to both no-activity and product control conditions (Fig.~\ref{fig:PC-640}D). Specifically, the partition coefficient for the active bath case is $2.3$ times higher than the no-activity control and $2$ times higher than the product control. The product control shows a slightly higher ($1.14$) partition coefficient compared to no-activity control (see Supp. Table~S1 for details).\\

\noindent To further verify these results, we perform bulk measurements by centrifuging the UBQLN2-450C system with the same conditions and quantifying the amount of protein in the dilute phase using spectrophotometry. Using this independent method, we find the amount of UBQLN2-450C in the dilute phase is much lower in the presence of the active bath than compared to the controls.  Specifically, the no-activity control is 5.7 times higher than active, while the product control is 3.4 times higher than active (Supp. Table~S1 for detail). These results are consistent with the concentration of UBQLN2-450C being higher in the condensed phase for the active system. Combining with previous results, this confirms that the active bath indeed drives more UBQLN2-450C into the condensed phase. 

\subsection*{Activity has little effect on enzyme partitioning or droplet material properties}
\noindent In addition to examining the partitioning of UBQL2-450C into the droplets, we can also quantify the partitioning of the urease enzyme (Supp. Fig. S1). We find that the urease concentration is slightly higher in the droplets for all cases with a partition coefficient of about 1.3. There is a slight increase in the partitioning to 1.37 $\pm$ 0.02 for the active bath case compared to the controls, which is significant implying that the activity could enhance the driving of all particles into the condensed state. \\

\noindent We also test the material properties of the droplets using fluorescence recovery after photobleaching (FRAP, Supp. Fig. S2). We find that the active case has a small, but significant decrease in the recovery time, implying that the mobility of the UBQL2-450C is higher in the droplets when there is an active bath compared to the controls. For all cases, the recovery is almost 100\%, but is slightly lower for the active case. This minor, but statistically significant change implies that the UBQLN2-450C molecules do not exchange with the background as quickly as the controls, implying increased interaction strength within the droplets. 

\subsection*{Simulations of sticker–spacer polymers in an active bath}
 
\noindent We investigate the role of activity in biomolecular condensation using coarse-grained molecular dynamics simulations of a “sticker–spacer” polymer system coupled to an active bath. Each polymer consists of alternating type A and type B stickers connected via implicit, stretchable spacer bonds. The polymers are modeled at a resolution in which each monomer has a radius of 1 nm, representing UBQLN-450C molecules. Specific interactions are implemented through short-ranged attractive potentials between unlike sticker types (A-B), reflecting a one-to-one binding affinity ($\epsilon_{AB}$). To avoid non-specific aggregation or polymer collapse, strong, short-ranged repulsive interactions are introduced between like sticker types (A-A and B-B), effectively preventing overlap beyond bound pairs (see Supplement for details)~\cite{zhang2024exchange}. The bath contains active particles of radius 4 nm, representing urease enzymes. The activity of these particles is increased by increasing their Pecl\`et number, $Pe$.  \\

\begin{figure*}[h!tbp]
    \centering
    \includegraphics[width=0.8\textwidth]{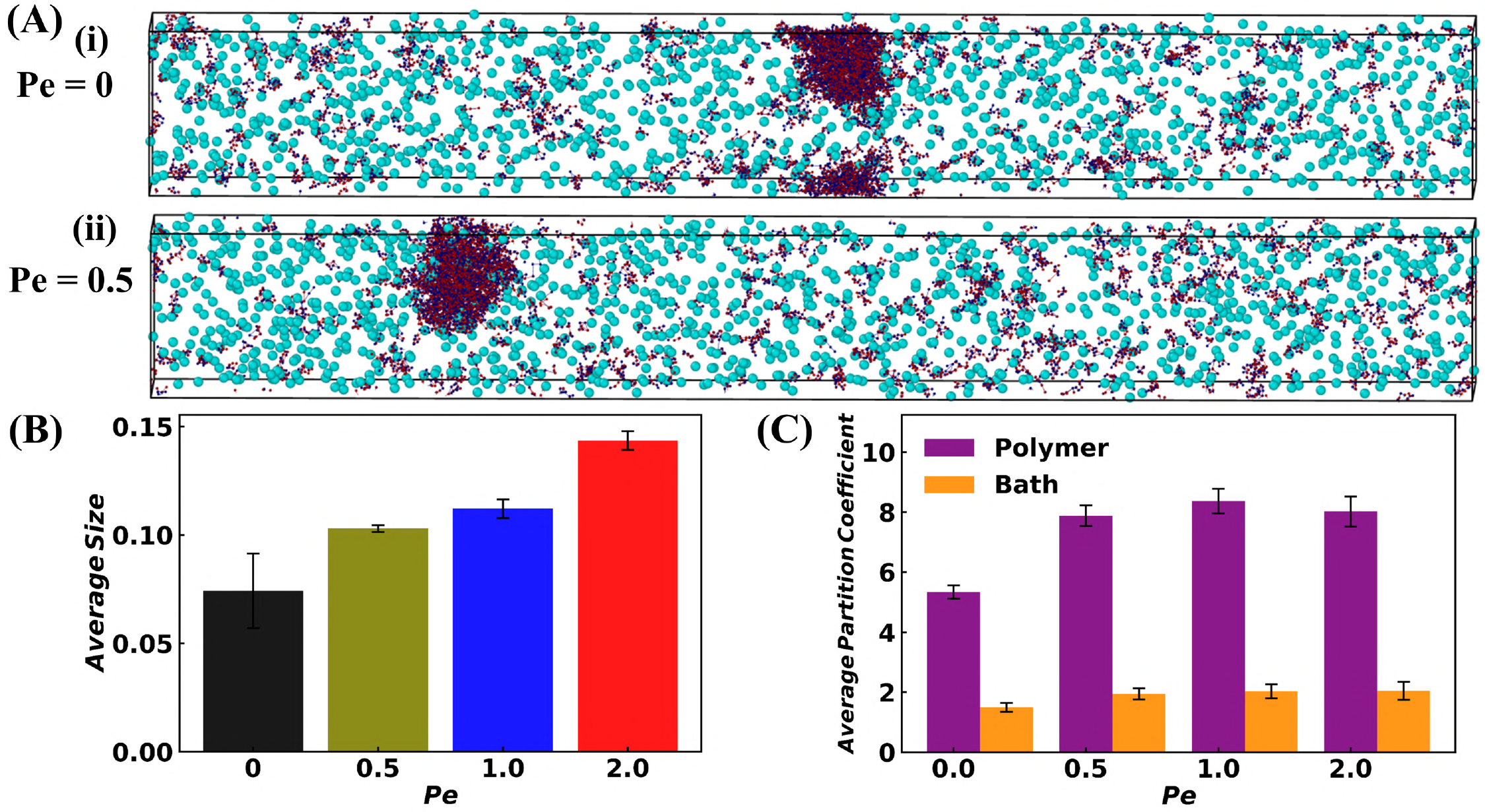}
    \caption{\textbf{Active bath controls condensation of sticker-spacer polymers.} (A) Snapshots of sticker–spacer polymers in (i) a passive bath ($Pe = 0$) and (ii) an active bath ($Pe = 0.5$). (B) Average size (fraction of particles) of the droplets. (C) Average partition coefficient for polymers and bath particles as a function of $Pe$, at fixed polymer binding affinity $\epsilon_{AB} = 5$.}
    \label{fig:simulation}
\end{figure*}

\noindent In our simulations, we observe the formation of droplet-like structures as a function of activity and binding affinity. Representative simulation snapshots (Fig.~\ref{fig:simulation}A) show that polymers self-assemble into droplet-like clusters in a passive bath ($Pe = 0$) and in the active bath ($Pe = 0.5$). In the active bath, the droplets also contain the bath particles. We characterize the size and composition of the droplet-like structures by computing the average size and the partition coefficients for both polymers and active bath particles. We designate that polymers and active bath particles are inside a droplet-like structure based on spatial proximity; namely, two stickers are considered connected, and thus part of the same droplet, if they lie within 1 nm of each other. The average size is defined as the fraction of polymer chains and active bath particles within a given droplet, normalized by the total number of particles in the system. We find that average droplet size increases monotonically with the active bath particle $Pe$ (Fig.~\ref{fig:simulation}B), indicating enhanced self-assembly in the presence of activity. For instance, the average cluster size increases by approximately 30\% from the passive case, $Pe = 0$, to $Pe = 0.5$, and continues to grow with increasing $Pe$. \\

\noindent To quantify species enrichment, we compute the partition coefficient, defined for each species as the ratio of its population in the dense phase to that in the dilute phase. Activity strongly enhances partitioning compared to the passive case. Notably, the polymer partition coefficient is approximately five times greater than that of the active bath particles, indicating preferential enrichment of polymers in the dense phase for all activity levels (Fig.~\ref{fig:simulation}C). In experiments, the difference was a factor of two (compare Fig. \ref{fig:PC-640}C to Supp. Fig. S1). The polymer partition coefficient increases by nearly 1.5-fold in the active bath ($Pe = 0.5$) compared to the passive case, with only a modest further increase at higher activity. In contrast, the partition coefficient of bath particles remains relatively low, increasing by just 15–20 \% from passive to $Pe = 0.5$, and showing minimal change at higher activity levels. These results are similar to what we observe in experiments, where we see a 7\% increase (Supp. Fig.~S1iv). \\

\noindent We also examine the polymer conformations in the dense and dilute phases using the radius of gyration ($R_g$, Supp. Fig. S4). When the binding affinity is low, ($\epsilon_{AB} = 4$), $R_g$ distributions are nearly identical across phases, regardless of activity (Supp. Fig.~S4 i,ii). When the binding affinity is high, polymers in the dense phase adopt more extended conformations than those in the dilute phase (Supp. Fig.~S4 iii,iv).

\section*{Discussion}
\noindent We combined an enzyme bath with a protein-based LLPS system to investigate how enzyme activity affects protein condensation. We specifically isolate the physical activity from the chemical through the use of a chemostatic chamber that can feed the enzymes and remove the products of the reaction. We observe that in the presence of the active bath, protein droplet size and partition coefficient both increase significantly at the same salt concentration and temperature. \\

\noindent Since UBQLN2-450C is known to condense more at higher temperature, these results suggest that the activity of the enzyme bath could be acting like an effective temperature to drive the protein condensation. Although this has been hypothesized and shown for micron-scale systems, this is the first demonstration of an active bath acting as an effective temperature for nanoscale objects. Further, this work implies that the background activity of the myriad of enzymes performing overlapping chemical reactions could also serve to mix the inside of the cell despite the complex physical nature of the cell interior. \\

\noindent Although slight, the partitioning of the enzyme increases and the material properties of the condensed phase of UBQLN2-450C are altered. The faster recovery time with slightly lower mobile fraction suggests the active bath is altering UBQLN2-450C interactions while increasing the mobility of mixing. We found quantitatively similar trends using a sticker-space polymer model with an active bath in which the bath particles were larger than the polymeric monomers given the relative size of the truncated protein and urease. The active bath particles corral the protein polymers and enhance their interactions while simultaneously increasing the $R_g$ of the polymers, indicating a dual role for the active bath. Interestingly, active LLPS theories indicate that activity does not simply make the system ``hotter'', but can also renormalize effective interactions between components~\cite{tjhung2018cluster,weber2019physics,Tayar2023}. \\

\noindent Recent studies of active bath systems examine the effects on phase separation, including \cite{Jambon-Puillet2024} that also use urease as the driver and \cite{Tayar2023} that uses a microtubule-based active fluid. Both of these experiments demonstrate that an active bath can alter the condensed state. However, in both of these systems, the condensed state is destroyed by the activity. Specifically, Jambon-Puillet, et al, show that the droplets swim toward their dissolution, dissolving themselves because the products are allowed to asymmetrically change the pH profile around the droplets \cite{Jambon-Puillet2024}. This is a chemically-induced driving, not the physically-induced driving we have performed here. Tayar, et al, covalently couple the microtubule-based driving to the condensed DNA nanostar system causing the condensates to stretch and be pulled apart like taffy \cite{Tayar2023}. While very exciting and demonstrating novel material properties, these studies are distinct from the results we demonstrate - namely that the physical nature of enhanced fluctuations from a bath of enzymes can control the phase of a protein to drive more phase separation, acting like an effective temperature. 

\section*{Materials and Methods}
\noindent For details, see supplemental. \\

{\bf Protein Reagents.} All proteins are diluted into low ionic strength sodium phosphate buffer (20 mM Sodium Phosphate, pH 6.83, 0.12M monobasic, 0.008M dibasic, pH adjust with NaOH). Urease is purchased from TCI with an activity level of 441 units/mg and used at a working concentration in all experiments at 100 units/ml. For experiments to visualize urease, proteins are labeled with Alexa-488 Fluor NHS ester using a commercially available kit (Fisher Scientific) and used with 6.5\% labeled protein. \\

\noindent Truncated Ubiquilin-2 (aa 450-624) is purified as previously reported \cite{Dao2018,Raymond-Smiedy2023} and used at a working concentration of $260 \mu M$. For experiments to visualize UBQN2-450C, proteins are labeled with Alexa-647 Fluor NHS ester using a commercially available kit (Fisher Scientific) and used with 4\% labeled protein. We add $300 mM$ NaCl to the sodium phosphate buffer to induce UBQN2-450C to phase separate. We keep the system at $37^{\circ}C$ for all microscopy experiments. \\ 

{\bf Experimental Chambers.} Chemostatically-controlled experimental chambers are created following the published protocol of \cite{Park2016} with the following changes. We modify the cleaning process, using Hellmanex III to clean the glasses for better results. We modify the amount of TMSPMA for the cover slip to 2\%. The 2\% allows F127 to stick to the surface of the chamber. Together, this process inhibits the phase separated proteins from binding to the cover slip. In addition, we use a reusable metal mask for the UV step with a slightly wider central channel (~1.8mm on mask). We also completely seal the cover slip microscope slide using epoxy.  \\   

{\bf Confocal imaging.} Condensates are imaged using spinning disc microscopy (Yokogawa CSU-W1) on an inverted Nikon Ti-E microscope with Perfect Focus and 100x oil immersion objective (1.49 NA) imaged onto a Andor Zyla CMOS camera. Images and image sequences are captured and saved as .nd2 files which are stacks of tifs with metadata. \\

{\bf Bulk measurment.} The UBQLN2-450C protein is mixed with urease and 1 mM urea (active bath), urease without urea (no-activity control), or  urease with 1 mM urea after the reaction is allowed to complete (products control). The samples are incubated at $37^{\circ}C$ for 10 minutes. To ensure the activity of active bath, urea was added at 8 minutes after incubation and right after incubation (but before centrifugation). The samples are centrifuged at 21000 $\times$ g for 30 seconds and the top layer of liquid is removed and measured using a Nanodrop at $\lambda = 280 nm$ to quantify the amount of protein. \\

{\bf Image analysis.} Images and image sequences are analyzed using ImageJ/FIJI, C++, and MatLab. ImageJ/FIJI was used to extract the size, droplet intensity, and background intensity for each z-level scan. This is done by using auto-thresholding (Intermode method for droplet size and intensity, MinError method for background intensity) to separate background and droplet. The background is selected by FIJI and the average intensity, for each z-level, would be measured. The droplet size and intensity are measured by  FIJI's particle analyzer. All the data are exported as .csv files. The maximum intensity (corresponding to the most focus z-level for each droplet) is found using C++ code. The size at the most focus z-level was used as the maximum droplet size in the analysis. The KS-test was perform by MatLab using kstest2 (two-sample Kolmogorov-Smirnov test). \\    

\section*{Acknowledgments}
\begin{acknowledgments}
\noindent We thank Carlos Casta\~{n}eda for providing the ubiquilin-2 construct and purification method and for helpful conversations and reading of this manuscript draft. This research is funded in part by the Alfred P. Sloan Foundation under grant G-2024-22546 to JLR and JMS. This work is funded in part from National Science Foundation grant from the Division of Materials Research CMP-2416012 to JLR and JMS. Additional funding was provided from Syracuse University.
\end{acknowledgments}

\section*{Data Availability}
\noindent Original data created for the study are available in a persistent repository upon publication at this DOI:

\section*{Supplemental Information}

\setcounter{figure}{0}
\setcounter{table}{0}

\titleformat{\section}
  {\normalfont\bfseries}         
  {\thesection}                  
  {0.75em}                       
  {\titlecap}    
\titlespacing*{\section}{0pt}{2ex plus .2ex minus .2ex}{1ex}
\titleformat{\subsection}[runin]
  {\normalfont\bfseries}
  {\thesubsection}
  {0.75em}
  {}
  { }                
\titlespacing*{\subsection}{0pt}{1.5ex plus .2ex}{0.5em}
\titleformat{\subsubsection}[runin]
  {\normalfont\bfseries\itshape}
  {\thesubsubsection}
  {0.75em}
  {}
  {}
\titlespacing*{\subsubsection}{0pt}{1.2ex plus .2ex}{0.5em}

\renewcommand{\figurename}{Fig.}
\renewcommand{\thefigure}{S\arabic{figure}}
\makeatletter
\renewcommand{\fnum@figure}{\textbf{\figurename~\thefigure}}
\makeatother
\renewcommand{\tablename}{Table}
\renewcommand{\thetable}{S\arabic{table}}
\makeatletter
\renewcommand{\fnum@table}{\textbf{\tablename~\thetable}}
\makeatother

\section{Additional Experiments}
\subsection{Urease partition coefficient is less sensitive to the activity.}
\begin{figure*}[h!tbp]
\centering
\includegraphics[width=0.95\textwidth]{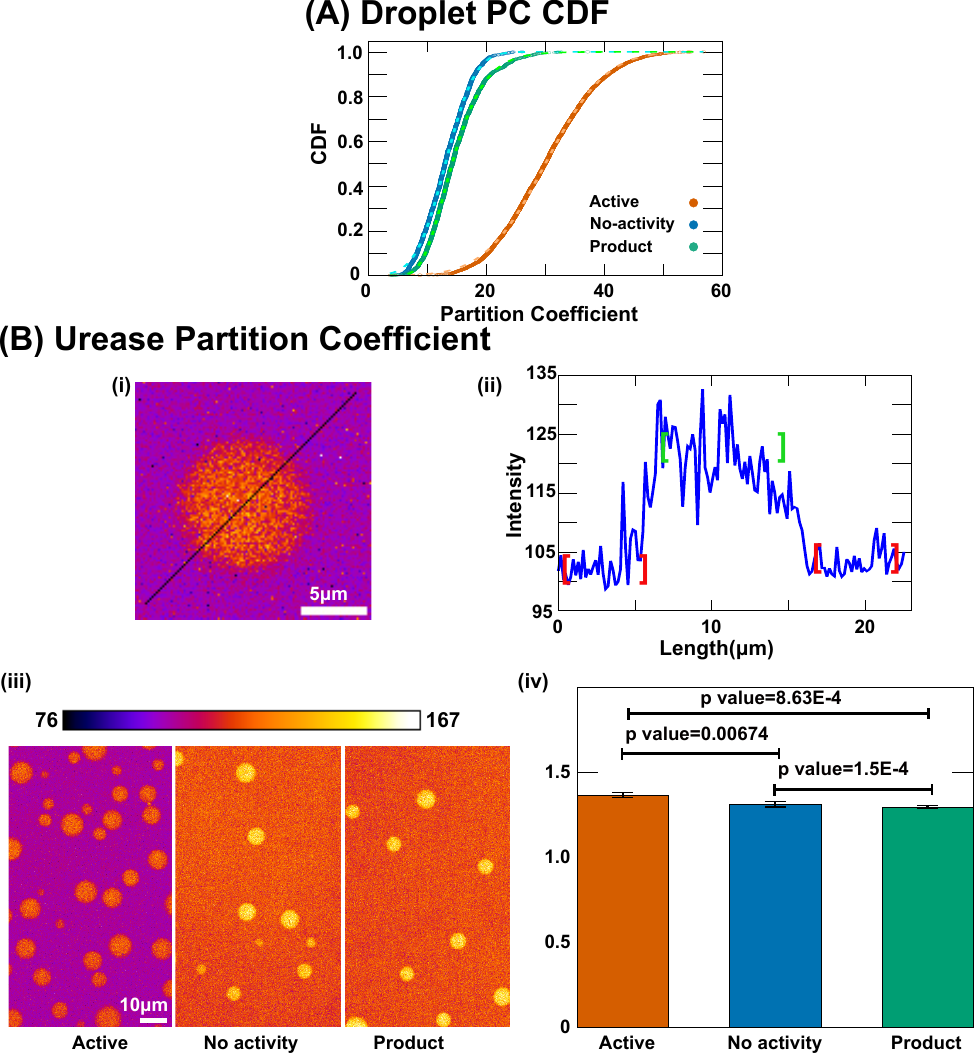}
\caption{\textbf{UBQL2-450C partition coefficient CDF and urease partition coefficient quantified for all conditions.} (A) CDFs of partition coefficient of droplets for active bath (orange filled circles), no activity control (blue filled circles), and product control (green filled circles). The CDF is fit with a Gaussian function for the active bath and no activity control or a log-normal function for the product control (dash lines). The means were used to determine the bar charts in Fig.~4C. (B) Urease partition coefficient. (i) The intensity profile is obtained from a line scan across the droplet (represented as black line) in a single z-plane. (ii) The intensity profile obtained from (i). Red bracketed region represents the data averaged to find the background intensity. Green bracketed region represents the date averaged to find the droplet intensity. (iii) Microscope images of droplets in the urease (488 nm) channel. Brightness represents the concentration of urease. All data on the same scale (scale bar 10 $\mu$m and same intensity look up table (given in the color scale bar from 76 to 167). (iv) Average urease partition coefficients obtained using the method shown in (i) for active condition (orange bar), no activity control (blue bar), and product control (green bar). Error bars represent SEM. Reported p-values calculated using the KS test. There are N = 105 data averaged in each condition. See Supp, Table.~\ref{tab:result} for values.}
\label{fig:PC-488}
\end{figure*}

In prior reports using an aqueous two-phase system (ATPS) of proteins condensed by polymers, the addition of urease enzymes showed high amounts of partitioning into the condensed protein phase causing large gradients in the products and pH across the boundary of the enzymes propelling the droplets to move \cite{Jambon-Puillet2024,Testa2021Sustained}. Given these prior results, we also investigate if the urease is more likely to be inside or outside of the condensed UBQLN2-450C droplets. Using a small fraction of labeled urease (6.5\% labeled), we quantified the partition coefficient by measuring the intensity of the cross-section of droplet images from a single z-plane (Supp. Fig.~\ref{fig:PC-488}Bi). We plotted cross-sectional intensity and determined the average intensity inside the droplet (Supp. Fig.~ \ref{fig:PC-488}Bii, green) and outside the droplet (Supp. Fig.~\ref{fig:PC-488}B, red) from the intensity profile. This method has been previously used \cite{Sahu2023} to compute the partition coefficient. \\

\noindent Comparing the images directly, it is qualitatively obvious that the active condition appears different from the no-activity and product control conditions (Supp. Fig.~ \ref{fig:PC-488}Biii) when imaged at the same magnification and intensity scale. Similar to the data for UBLQN-450C, the intensities in the droplets and outside are shifted, most likely becasue there are more droplets. Yet, the ratios of the internal to external intensities (the partition coefficients) are similar for all three conditions (Supp. Table \ref{tab:result}, Supp. Fig.~\ref{fig:PC-488}Biii,iv). Specifically, the active case has a partition coefficient of 1.37 $\pm$ 0.02 implying there is 37\% more urease in the droplet compared to the background. This is very low, especially compared to the amount of UBQLN2-450C in the droplets ($\approx$ 30 $\times$, or 3,000\% increase, Supp. Table \ref{tab:result}, Fig.~4C). Comparing the amount of urease in the droplets between the active case and the no-activity control or the products control, we see both controls have a about a 30\% increase (Supp. Table \ref{tab:result}, Supp. Fig.~\ref{fig:PC-488}Biii,iv). \\

\noindent Even though the activity of urease does appear to drive more enzymes into the droplets, the increase is small, especially compared to previous results using the APTES condensed system where the partioning was described as "strong" or infinite \cite{Jambon-Puillet2024,Testa2021Sustained}.  Combined with the fact that we are using $1~mM$ urea, we do not believe that the local pH change is responsible for the phase diagram alteration we observe. Further, the product control condition has the same pH shift, since that is the case where the reaction runs to completion and has the carbon dioxide and ammonia in solution, the cause of the increased pH. 

\subsection*{Active bath affects material properties of LLPS.}
\begin{figure*}[h!tbp]
\centering
\includegraphics[width=0.95\textwidth]{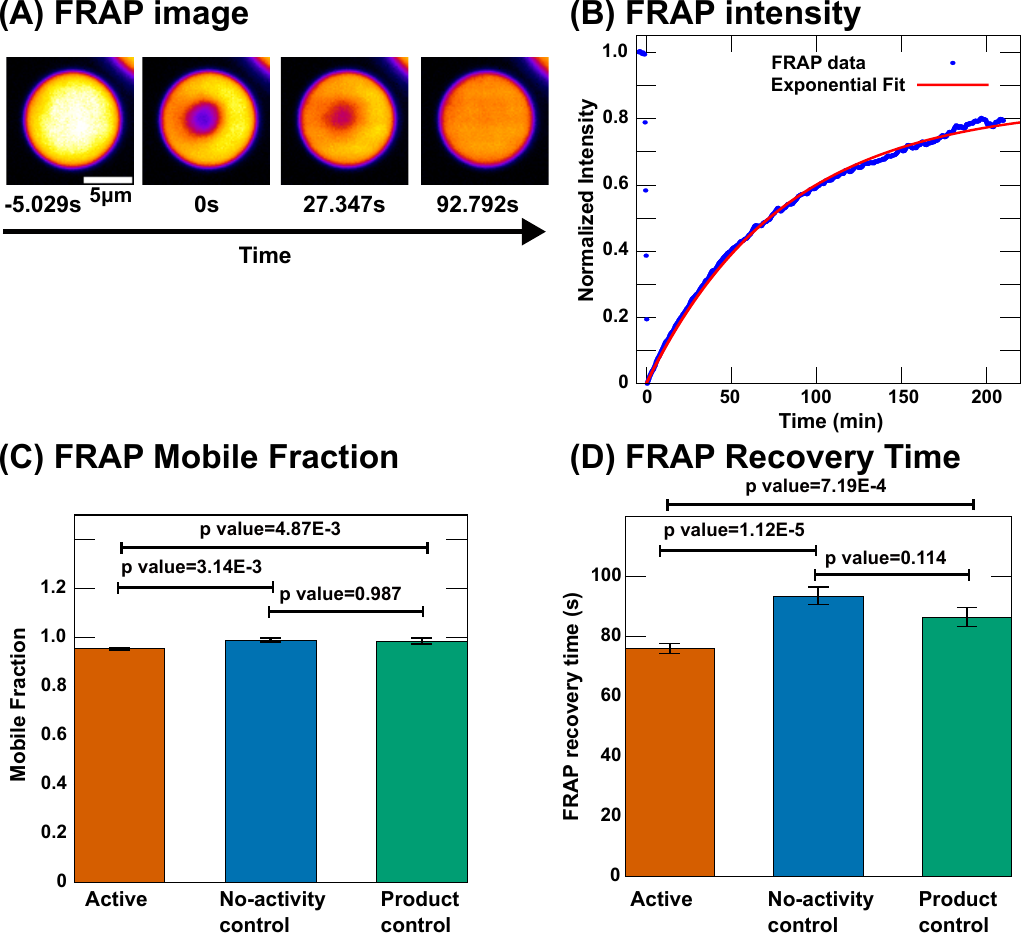}
\caption{\textbf{UBQL2-450C FRAP mobile fraction and recovery time quantified for all conditions.}(A)Timeseries of FRAP on a single droplet with 0s as the time just after the photobleaching using a 405 nm laser. The droplet undergoes global photobleaching over time hence the intensity of droplet at the end is less than at the beginning. (B) Droplet intensity over time after photobleaching. We fit an exponential function (Eqn.~\ref{eqn:exp}) and quantify the mobile fraction and recovery time from the fit parameters. (C) The average mobile fraction obtained using Eqn.\ref{eqn:exp} for active condition (orange bar, N = 68), no activity control (blue bar, N = 3), and product control (green bar, N = 42). Error bars represent SEM. Reported p-values calculated using the KS test. (D) Average FRAP recovery time obtained using same method in (C) for active condition (orange bar, N = 68), no activity control (blue bar, N = 3), and product control (green bar, N = 42). Error bars represent SEM. Reported p-values calculated using the KS test.}
\label{fig:FRAP}
\end{figure*}

Given how the active bath can drive more UBQLN2-450C molecules into the dense phase, we want to know if this activity can alter droplet dynamics as well. To investigate the dynamics of UBQLN2-450C molecules, we perform fluorescence recovery after photobleaching (FRAP) on droplets (see Supp. Fig.~\ref{fig:FRAP}A,B). We observe that all the mobile fractions are close to 1 (100\% recovery, Supp. Fig.~\ref{fig:FRAP}C). In the active case, droplets have a slightly lower mobile fraction ($0.953 \pm 0.005$), compared to no activity control ($0.989 \pm 0.008$) and the product control ($0.98 \pm 0.01$). This implies that the immobile portion within droplets are increased by the active bath, suggesting some changes in the multivalance interaction between the UBQLN2-450C polymers. We note that the difference between the active case and the control experiments, although statistically significant, is very small compared to the changes in droplet size and partition coefficient (Fig.~3, 4).\\

\noindent We also observed the FRAP recovery time decreases (faster recovery) with active bath compare to no activity and product controls (see Supp. Table. \ref{tab:result}, Supp. Fig.~\ref{fig:FRAP}D). In comparison, the recovery time with active bath is about 0.81 times lower than no-activity control and about 0.88 times lower than product control. While the product control is slightly faster compared to no activity control (0.98 times faster). This implies that the active bath may be able to increase the diffusion within droplet. \\

\begin{table*}[h!t]
\centering
\caption{Results from analysis}
\label{tab:result}
\begin{tabular}{lccc}
Analysis & Active Bath & No-activity control & Product control  \\
\hline
Average droplet size($\mu m^2)$ & $45.1\pm0.6$ & $23.5\pm0.6$ & $13.4\pm0.4$\\
Droplet density & $0.0031+0.0004$ & $0.0015+0.0002$ & $0.0019+0.0002$\\
Area fraction & $0.139,0.161,0.137$ & $0.035,0.040,0.034$ & $0.026,0.03,0.025$\\
Average partition coefficient of UBQLN2 & $30.0\pm0.1$ & $13.1\pm0.1$ & $14.9\pm0.1$\\
Bulk measurement average (A.U.) & $0.70\pm0.09$  & $3.96\pm0.1$  & $2.3\pm0.3$ \\
Average partition coefficient of urease & $1.37\pm0.02$ & $1.31\pm0.0.02$ & $1.30\pm0.01$\\
FRAP mobile fraction & $0.953\pm0.005$ & $0.989\pm0.008$  & $0.98\pm0.01$\\
FRAP recovery time(s)& $76\pm2$ & $94\pm3$  & $86\pm 3$\\
\hline
\end{tabular}
\vspace{0.5ex}
\par\footnotesize{Results are shown as mean $\pm$ SEM except for droplet density and area fraction. For density, the error is the percentage of missed droplet (see method-Analysis-Droplet counting restriction) .For area fraction the results are shown as mean,max value,min value}
\end{table*}

\section*{Detailed Materials and Methods}

\subsection*{UBQLN2-450C protein purification.} We use a truncated ubiquilin-2 with 450 - 624 aa as a model LLPS system, denoted UBQLN2-450C. This protocol was modified from \cite{Dao2018}.

\subsubsection*{Protein expression.}
 The UBQLN2 plasmid (Addgene 176852) is expressed using T7 Express lysY competent E. coli (New England Biolab, C3010) on LB plate (50 $\mu$g/mL Kanamycin and 35 $\mu$g/mL Chloramphenicol). The liquid culture is incubated in shaking incubator at $37^{\circ}$C until OD$_{600}$ reaches between 0.6 to 0.8. Isopropyl $\beta$-D-1-thiogalactopyranoside (IPTG) is added  to a final concentration of 0.05\%(v/v), the liquid culture is allowed to incubate overnight. 
 
\subsubsection*{Protein purification.}
We centrifuge the liquid culture at 4000 $\times$g for 15 minutes at $4^{\circ}$C. The pellet is frozen at $-80^{\circ}$C for at least one hour before adding lysis buffer (0.25 mM Phenylmethanesulfonyl fluoride (PMSF), 0.99 mM Ethylenediaminetetraacetic acid (EDTA) at pH 8, 49.71 mM Tris at pH 8, 2.49 mM MgCl$_2$, and 0.0746\% (v/v) DNase I) during the thaw. The resuspended pellet is sonicated and centrifuged at 20,000 $\times$ g for 20 minutes at $4^{\circ}$C. The supernatant is placed in $37^{\circ}$C water bath for 10 minutes before 500 mM NaCl (final concentration) was added. The NaCl causes the UBQLN2-450C protein to phase separate, making the solution turbid. We centrifuge the solution at 5000 $\times$g for 15 minutes at room temperature to collect the pellet. We dissolve the pellet in sodium phosphate (20 mM Sodium Phosphate, pH 6.83, 0.12M monobasic, 0.008M dibasic, pH adjust with NaOH). We place the protein on ice for 4-6 hours to let the pellets dissolve. We mix the pellet well and add 500 mM NaCl and centrifuge again to collect the pellet. We dissolve the second pellet with sodium phosphate buffer on ice for 4 to 6 hours. We centrifuge the resuspended pellet at 21,000 $\times$g for 3 minutes at $4^{\circ}$C to remove any protein that cannot be resuspended. 

\subsubsection*{Protein desalting and labeling.}
We use a PD-10 desalting column, G-25M (Cytiva, 17085101) for removing the NaCl from the protein solution. For full protocol see the user manual. Briefly, we perform two desalting steps, gravity and spin. The gravity protocol was performed first and the then spin protocol with a new desalting column. We use sodium phosphate buffer as the equilibration buffer in the protocol. Once the desalting is complete, the protein solution was centrifuged at 21,000 $\times$g for 3 minutes at $4^{\circ}$C to further remove any debris. Protein solutions are aliquoted and drop frozen by liquid nitrogen, and stored at $-80^{\circ}$C for long term storage. To observed the droplets under microscope, we label the purified UBQLN2-450C proteins with labeling kit Alexa Fluor NHS ester 647 following the directions from the manufacturer. 

\subsection*{Enzymes.}
We use the enzyme, urease, as the model active bath (see Fig.~1). Urease (TCI-U0017) and urea (VWR-BDH4602) are prepared fresh each week and dissolved in sodium phsophate buffer. Urease stocks are stored at $4^{\circ}C$ and urea stock are stored at room temperature.  To observed the urease partitioning into droplets using fluorescence microscopy, we label the purified UBQLN2-450C proteins with labeling kit Alexa Fluor NHS ester 488 following the directions from the manufacturer. \\

\noindent The urease activity is tested using phenol red. Phenol red is a pH indicator that changes color from yellow to pink within pH 6.8 to 8.4 \cite{Kuan2014}. The products of urea, ammonia and CO$_2$, in combination, both increase the pH of the solution, turning phenol red to pink from yellow. To perform the test, we first add the urea into the diluted phenol red. No color change should occur, as a change of color from yellow to pink would indicate that the urea is already hydrolyzed. We then add urease into the mixture, making the final solution of 28$\mu$M phenol red, 16$\frac{unit}{ml}$ urease, and 2 mM urea. We shake the tube few times to mix the urease, and a pink color should appear, indicating that urease can catalyze the urea hydrolysis reaction. Note that the phenol red solution should be diluted in ddH$_2$O, since this test is based on pH changes by products. Using a large amount of buffer would hinder this change and possibly no color change would be observed, unless large a amount of urea is used.

\subsection*{Microfluidic chamber protocol.} 
There are two issues with using urease as active bath: 1) it requires urea, and urea hydrolysis is catalyzed by urease, hence it needs a source of urea to maintain the activity. 2) as urea gets hydrolyzed, it forms two products, $CO_2$ and ammonia, which can raise the pH levels and alter the ions in solution. This can alter the behavior of LLPS due to chemistry, not physics. To resolve those, we use a microfluidic chamber with the ability to add urea and remove products. The chamber consists of three lanes separated by two semi-permeable membranes made by polyethylene glycol diacrylate (PEGDA) 400 (see Fig.~2A). The PEGDA-400 allows small molecules such as NaCl and urea to pass through while trapping larger molecules (proteins). The microfluidic chamber is required to constantly supply enzymes with substrate and remove the products to create a chemostatic environment for testing. The chamber design is described more in the main text. Here, we describe the procedure to create the chambers.

\subsubsection*{Cleaning and glass treatment.} 
Cover slips (24 $\times$ 60 mm, No.1) and microscope slides (25 $\times$ 75 $\times$ 1 mm) are sonicated in a 1\% Hellmanex III solution for one hour. The glasses are then rinsed with ddH$_2$O, acetone, isopropanol, and ddH$_2$O again. The glass is dried by air and further cleaned by a UVO machine for 20 min. Two 3-(Trimethoxysilyl)propyl methacrylate (TMSPMA, Sigma M6514, 248.35 g/mol) solutions (2\% and 0.5\%) are prepared in isopropanol. The 2\% TMSPMA was dripped onto the cover slips, while 0.5\% was dripped onto the slides. The glasses are then baked in a $70^{\circ}C$ oven for 90 minutes. 

\subsubsection*{Chamber assembly.} 
Once the glass is baked, we combine the two glasses in a cross shape with the cover slip on the bottom and using two parafilm strips cut to 3 cm $\times$ 1 cm as spacers between the cover slip and slide, as diagrammed (Fig.~2A). The chamber chamber edges are sealed using a soldering iron to melt the parafilm. At this point, the assembled chambers can be stored with desiccant in a box to keep moisture low.

\subsubsection*{Semipermeable membranes.}
To create the semipermeable membrane walls inside the chamber, in situ, we use a photocrosslinkable polymer. We mix 2 $\mu$L of undiluted 2-hydroxy-2-methylpropiophenone (photo-crosslinker, Sigma-Aldrich 405655) with 120 $\mu$L of undiluted poly(ethylene glycol) diacrylate 400 (PEGDA-400, Polysciences-01871-250 with specific gravity of 1.1080 - 1.1310 g/mL); this is solution A. We mix 0.1 mM Tris base with solution A at a 5:1 ratio. Specifically, $320\mu$L Tris Base with $80\mu$L solution A. This membrane solution is then kept on ice for 10 minutes before use.\\

\noindent Using a test chamber (the same chamber without any cleaning or surface treatment), we tape the chamber (cover slip side) to a metal mask with 0.5 mm slits spaced 1.8 mm apart. We inject enough membrane solution to fill the test chamber, around 50 $\mu$L. We place the chamber and mask on a UV light box with the light incident from cover slip side for about 12 seconds. The time was determined empirically and depends on the UV light box, the placement on the box, and the distance from the lamp. When the semipermeable membranes form, they appear as two opaque, white lines inside the chamber. We next repeat the process with a surface-treated chamber so that the semipermeable membrane will crosslink to the cover slip and slide. We ensure that the membranes form to the ends of the chamber to avoid leakage at the entrances by creating a puddle of membrane solution at both entrances of the chamber. We seal the edges between the cover slip and slide using two additional pieces of parafilm (3 cm $\times$ 1 cm) and adding a small amount of epoxy on them. We glue the parafilm on the edge between the cover slip and slide to eliminate possible leakages.

\subsubsection*{Sample and tube insertion.}
Once the semipermeable membrane forms, we flow 250 $\mu$L of Pluronic-F127 (1.25\% w/v) into the chamber. We allow the Pluronic-F127 to incubate for at least 5 minutes. To keep chamber hydrated, we leave chamber in a closed box with damp kimwipes. Meanwhile, we prepare four wide tubes (Tygon outer diameter of 0.030 inches, inner diameter of 0.01 inches, ND-100-80) by flowing sodium phosphate buffer through them by submerging one side of the tube to a reservoir with buffer. The tube length depends on the microscope setup, and we make it as short as possible. For each wide tube, we insert a thin tube (UDEL capillary tubes from Incom, outer diameter of 95 $\mu$m, inner diameter of 74 $\mu$m) into the side of the Tygon tubes without the reservoir. Each thin, UDEL tube is around 0.8 inches in length. When all the wide Tygon tubes are prepared, we flow $70\mu$L protein sample into the chamber. We then insert the free end of the thin, UDEL tubes into one of the outer lanes of the chamber. Once the thin tubes are in, we secure the tube combination to the chamber using tape. We repeat this for all four inlet and outlet for the outer lanes of the chamber.

\subsubsection*{Seal chamber and storage.}
Once all tube combinations are in and the sample is inserted, we seal the two open sides with epoxy. We leave the chamber at room temperature for 30 minutes to let the epoxy cure, covering the chamber with aluminum foil if the sample is light sensitive, making sure that the aluminum foil does not touch epoxy. We then move the chamber with the buffer reservoirs to a $4^\circ$C refrigerator overnight, or leave it for another 2.5 hours at room temperature, before use. 

\subsection*{Experimental design.}

\subsubsection*{Sample preparation.}
All experiments are performed using 260 $\mu M$ UBQLN2-450C with 4\% labeled protein, 100 $\frac{unit}{ml}$ urease with 6.5\% labeling, and 300 mM NaCl to induce UBQLN2-450C to phase separate. We also kept the system at $37^{\circ}C$ for all microscopy experiments using an enviromnmental chamber. To maintain the pH at 6.83, we used a 20 mM sodium phosphate buffer (0.12 M monobasic, 0.008 M dibasic, pH adjusted with NaOH). To minimize the affect of products, we use 1 mM urea to activate the activity of urease. However, due to urea hydrolysis (see Fig.~1Bii), we also need an abundant source of urea to keep the bath active, hence all experiments with active bath is done with the microfludic chamber (see Microfluidic Chamber section).\\

\noindent Before use, we thaw and centrifuge at 20000 $\times$g for 3 minutes at $4^{\circ}C$. We only used the supernatant to avoid insoluble aggregates. We then remeasured the centrifuged the concentration of UBQLN2-450C using a Nanodrop spectrophotometer and the extinction coefficient of 5500$M^{-1}cm^{-1}$ \cite{Dao2019}. Once this is done, we prepare the chamber and leave the proteins on ice before use.

\subsubsection*{Control experiments.}
As described in the main text, we perform two different control experiments to compare to the active bath experiments. The negative control has urease present without urea to examine the effects of the protein without the activity. The second positive control combines the urease with the products from the catalysis reaction to examine the effects of the chemistry alone without activity. \\

\noindent For the  active bath, a solution of 260 $\mu M$ UBQLN2 (with 4\% labeling), 100 $\frac{unit}{ml}$ urease (6.5\% labeling) is made. For controls, the same solution is made as for the active bath, with additional 300 mM NaCl. For the product control, the same protein solutions are made with 300 mM Nacl and products made from 1 mM of urea being hydrolyzed by urease. The product was made by mixing urea and urease and incubated for at lease 4 hours at $37^{\circ}C$ to ensure all urea were turned into byproduct. Similar to urease, a phenol red test is done to ensure the byproduct is form before use. For both controls, the solutions are incubated at $37^{\circ}C$ for 10 minutes before the sample is flowed through to induce phase separation. \\

\noindent The control experiments do not require constant exchange of small molecules, so all control experiments were performed with chambers without semipermeable membranes. These chambers consist a silanized cover slip (22 $\times$ 30 mm; Thermo Fisher Scientific) and an untreated microscope slide separated by two double-sided tape spacers. The cover slip is treated with Repel Silane (Cytiva 17-1332-01) using previously published protocols \cite{Dixit2010,Chauhan2024}. Briefly, we silanize by first cleaning the cover slip by UVO machine for 20 minutes, then wash with acetone for 1 hour, ethonal  for 10 minutes, and 0.1 M KOH for 15 minutes. We submerge the cover slip with ddH$_2$O 3 times between each chemical wash for 5 minutes each. After washing, the cover slip is air dried, usually dry overnight in a fume hood, we submerge the cover slip with Repel Silane for 5 minutes. We then rinse the cover slip by submerging it in ethonal  for 15 minutes and ddH$_2$O 3 times for 5 minutes each. Note that this silanization would hold up for about a month. The chambers are coated with 1.25\% Pluronic-F127 to coat the surfaces. The sample is pipetted into the flow path between the tape strips and the open ends are sealed with epoxy prior to visualization on the microscope. 

\subsubsection*{Spinning Disc Microscopy.}
Condensates are imaged using spinning disc microscopy (Yokogawa CSU-W1) on an inverted Nikon Ti-E microscope with Perfect Focus and 100x oil immersion objective (1.49 NA) imaged onto a Andor Zyla CMOS camera. Images and image sequences are captured and saved as .nd2 files which are stacks of tifs with metadata.\\

\noindent We took two types of data using the spinning disc microscope: Time scan of 210 seconds each with fluorescence recovery after photobleaching (FRAP, done by a $\lambda=405$ nm laser, see Supp. Fig.~\ref{fig:FRAP}A) and z-level scans taken with two wavelengths, $\lambda = 640$ nm at 38\% power to image the UBQLN2-450C and $\lambda = 488$ nm at 45\% power to image the urease. The z-level scan has increment of $\Delta z=0.3\mu m$. \\

\noindent Due to the possibility of the system still equilibrating due to droplet coalescence and Ostwald ripening, all data are taken around the same time for each experiment for comparison. Specifically, all FRAP data are taken between 115-132 minutes after phase separation is induced at 300 mM NaCl and $37^{\circ}C$, and z-level scan data are taken between 137-155 min after phase separation is induced at at 300 mM NaCl and $37^{\circ}C$. \\

\noindent Photobleaching was performed using a 405 nm laser with a circular spot to irreversibly deactivate the fluorescent dye. Once the dye is deactivated, we took a 210-second video to see the recovery of dye only in the UBQLN2 channel, (Supp. Fig.~\ref{fig:FRAP}A). All data are taken between 115-132 minutes (FRAP) and 137-155 minutes (z-level scan) after phase separation is inducted. Each experiment is performed three times, in three different chambers. \\

\subsubsection*{Bulk measurements.}
We also measure the dilute phase of each experiment, the active bath, negative and positive controls, using the Nanodrop spectrophotometer. We first make all three experiments in tubes simultaneously. Each sample is kept on ice to inhibit phase separation. For the active bath case, we do not add urea at first because it would run out very fast. For the product control, we do not add in the product at first, and for no activity control, we do not add all the buffer in to maintain the same concentration between experimental set. We then incubated all three tubes for 8 minutes, at $37^{\circ}C$ to induce the phase separation. At the end of the 8 minutes, the missing chemicals are added to each; specifically, the active bath adds urea to make 1 mM urea, the positive control with product adds the urea products to make 1 mM urea products, and the no activity control adds the same amount of buffer to maintain the same concentration compared to the other two. After addition, each was incubated for two more minutes. Next, we add a second round of small molecules that are the same as the first round to maintain the same concentration. Finally, we centrifuge all three samples at 21,000 $\times$g for 30 seconds to concentrate all the droplets at the bottom of the tube. We removed the top layer of liquid, which is the dilute phase and measure the concentration with the Nanodrop at $\lambda = 280$ nm. Similarly to the microscopy experiment, each experiment were performed three times.

\subsection*{Image analysis.}
ImageJ/FIJI is used to analyze the microscope images imported as Nikon .nd2 files which are stacks of tiff images with metadata attached using the Bioformats Importer plugin. 

\subsubsection*{Droplet detection.}
The following FIJI detection method is adopted from Mahdi-Moosa's github (https://github.com/Mahdi-Moosa/PartitionCoefficient-ImageJ\_Macro). We used FIJI's automatic thresholding method called Intermodes to create a mask of each slice of the z-stack of images of droplets. From that mask, we used the Analyze Particles function to identify droplets where we had a circularity cut off of 0.85 and was at least 10 pixels to be considered a droplet. \\

\noindent Above, we use several methods to remove droplets that are not fully imaged from our quantification. First, we reject droplets that are not circular. Second, we do not count the holes within a droplet as a separate droplet. Third, we exclude the droplet that merges, as that would cause a confusion in droplet counting. Last, we exclude the edge droplets as FIJI's Analyze Particles allows us to remove droplets on the edge. We still found that some parts of droplets appeared away from the edge, but eventaully overlapped the edge at a different z-level. To remove the droplets that overlapped the edge in a higher z-level, we limited the observed area when finding the maximum brightness to exclude particles with their centers within 45 pixels of the edge of the frame. Essentially, we only count the droplets that are within a box of $2003 \, by \, 2003$ pixels.

\subsubsection*{Droplet size measurement.}
\begin{figure*}[h!tbp]
\centering
\includegraphics[width=0.95\textwidth]{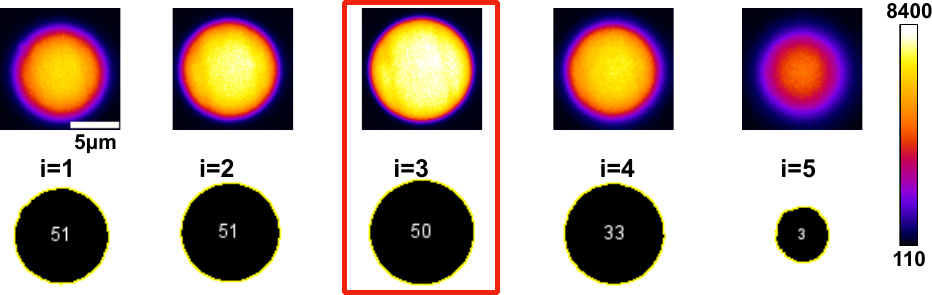}
\caption{\textbf{Example of droplet size selection using FIJI measurement.} (Top) Example z-slices of a droplet showing the original intensity using the fire look-up table set to 110 - 8400 AU.  (Bottom) Detected droplet region (black) using FIJI Analyze Particles. The number inside the detected region is the order of which FIJI detects the droplet at a specific z-slice, which we do not use. For each successive z-slice, the difference between the center is calculated ($\Delta r(x_i,y_i) = \sqrt{(x_i-x_{i-1})^2+(y_i-y_{i-1})^2}$ and each successive z-slice is considered the same droplet only if $\Delta r(x_i,y_i)\leq15$ pixels. The C++ program would select out the maximum intensity z-slice (red box) among the z-slices and the corresponding droplet area would be used as the maximum droplet size.}
\label{fig:DP_select}
\end{figure*}

Using the image masks created from thresholding, the regions of interest that correspond to each droplet on each z-slice were superimposed on the original z-stack. Using these regions of interest, we measured the x-y position of the center, the area, the intensity, and the circularity of each droplet for each slice using FIJI Analyze Particles. We considered the regions from different z-slices to describe the same droplet if the x-y center was within 15 pixels for all z-slices. For each droplet, we identify the vertical center of the droplet by finding the z-slice where the maximum intensity occurs. The maximum area of each droplet is extracted from this z-slice (see Supp. Fig.~\ref{fig:DP_select} for example). 

\subsubsection*{Droplet intensity measurement.}
As described above, the droplet intensity was measured to determine the droplet size. That same maximum intensity value was used for the partition coefficient measurement to quantify the amount of UBQLN2-450C inside that specific droplet. The average intensity of the droplet is used as the numerator in the partition coefficient ratio. 

\subsubsection*{Background intensity measurement.}
To calculate the partition coefficient, we also need to measure the background intensity of the UBQLN2-450C. This was performed for the entire background in a specific field of droplets. In this case, we use FIJI's built-in MinError method to threshold the image. This separates the background and foreground (i.e., the droplets) with gray-values of 0 and 255, respectively. Then the two gray values switch by inverting the image such that the droplets are zero and the background is 255. The region of interest that encompasses the pixels that are at 255 (the background) are selected. The same region is restored onto the original image (without thresholding) and the average intensity of the background is quantified. Thus, obtaining an average background measurement, which is the denominator of the partition coefficient ratio.

\subsubsection*{Urease intensity measurement in droplets.}
Due to the low partitioning of urease resulting in a partition coefficient close to one, the auto thresholding methods cannot be used. Instead, we manually select a frame through the middle of the droplet. We draw a line through the center of the droplet and use it to create an intensity profile (Supp. Fig.~\ref{fig:PC-488}A,B). The average intensity inside and outside of the droplets are measured by averaging over many points and the partition coefficient is determined by the ratio of the intensity averages both inside and outside the droplet.

\subsubsection*{Droplet counting.}
Due to the restrictions we imposed on detecting our droplets including circularity, object pixel limitation, FIJI's edge exclusion, merging exclusion, and our more-restrictive box edge exclusion, the number of droplets we count is under-counted because these droplets are in the area, but we exclude them from analysis. It could also be possible that droplets initially outside of the box could diffuse into the box during our z-level scan, but this rarely happens because each z-scan is done at least 2 hours after droplets are induced and thus all droplets are settled at the bottom and not floating in solution. When this occurs, we also exclude the droplet to avoid confusion in counting. \\

\noindent In addition, we notice that some droplets, especially smaller, less bright droplets, are missed because of the threshold settings. When small and large droplets are in the same frame, the larger droplets appear brighter and dominate the threshold settings. This causes the smaller droplets to be excluded from the counting. One example is Fig.~4Biv, droplet number 4, the detected radius is slightly smaller than the droplet. \\

\noindent In all these cases, we as systematically under-counting because we do not include these objects are droplets, even though they are. Conversely, over-counting means objects detected by FIJI that are not droplets, but count as droplets. From visual inspection of our analysis, we do not appear to have over-counting, mainly due to the circularity lower bound of 0.85, which seems sufficient enough to exclude anything that is non-circular. To estimate the systematic under-counting error, we identified the droplets that were excluded, mainly due the box-edge exclusion, and find the percentage of missed droplets with respect to the counted droplet. This difference is represented as the upper error bar in Fig.~3E,F of the main document.

\subsection*{Area fraction calculation.}
To calculate the area fraction occupied by droplets, we use two different approaches that were basically equivalent. First, we use 
\begin{align}
Area\, fraction &= \frac{\#\,of\,droplet*average\,size}{Total\,observe\, area} \\
                &= dropelt\,density*average\,size   
\label{eqn:AF} 
\end{align}
to calculate the area fraction for each case. We can also calculate the area fraction by summing up all droplets' area:
\begin{align}
Area\, fraction = \frac{\sum_{i=1}^{N}area_{i}}{Total\,observe\, area} 
\label{eqn:AF_sum} 
\end{align}
where $N$ represents the total number of droplets for each case (N=2933, 1404, and 1625 for active bath, no-activity control, and product control). Both methods yield similar results, see table.~\ref{tab:AF} below. These results are both within error due to the larger systematic error of under-counting, as described above.

\begin{table}[h!]
\centering
\caption{Area fraction by average area (Eqn.\ref{eqn:AF}) method vs summation method (Eqn.\ref{eqn:AF_sum})}
\label{tab:AF}
\begin{tabular}{lrrr}
Case & Average area method & Summation method \\
\midrule
Active Bath & 0.1393 & 0.1312  \\
No-activity & 0.0347 & 0.0313  \\
Product control & 0.0257 & 0.0230 \\
\bottomrule
\end{tabular}
\end{table}

\subsubsection*{FRAP.}
Using FIJI, we manually select the area that was photobleached by 405 nm laser using a circular region of interest and obtain the intensity of the selected area over the 210 seconds (see Supp. Fig.~\ref{fig:FRAP}A,B). Besides the photobleaching caused by 405 nm laser, there is also a global photobleaching caused by the 640 nm laser for imaging. To address the global photobleaching, we selected a neighboring droplet, that is not photobleached by 405 nm laser, as a reference to correct the global photobleaching. This method is presented in \cite{FRAP2021}. We fit the results with an exponential function:

\begin{align}
Int(t) = A(1-e^{\frac{-t}{\tau}})  
\label{eqn:exp}
\end{align}

\noindent
where $A$ is the mobile fraction and $\tau$ is the recovery time (see Supp. Fig.~\ref{fig:FRAP}B). 

\section*{Simulation Details}
\begin{figure*}[h!]
    \centering
    \includegraphics[width=0.8\textwidth]{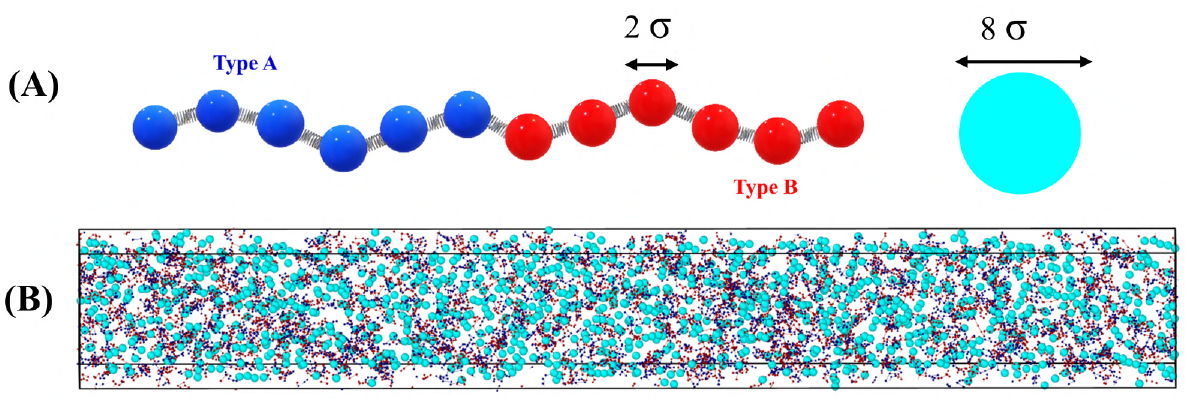}
    \caption{\textbf{Sticker-spacer polymers in active bath.} (A) Schematic of a sticker–spacer polymer (stickers: red/blue; spacers: springs) and a bath particle. Polymer monomer diameter = 2 $\sigma$ and the bath particle diameter is 8 $\sigma$, where $\sigma = 1$ nm. (B) Representative simulation snapshot of sticker–spacer polymers in an active bath for Pe = 0.5 and binding affinity $\epsilon_{AB} = 5$ $k_BT$, captured at simulation time $\tau_{\mathrm{sim}} = 1000 \, \tau$, at an earlier simulation time.}
    \label{fig:model_sim}
\end{figure*}

We perform coarse-grained molecular dynamics simulations using LAMMPS ~\cite{plimpton1995fast} to investigate the physics governing condensate formation and the influence of an active bath on dense and dilute phases. Polymers are modeled as linear chains of spherical monomers with radius $r_p = 1$ nm, connected by springs. The active bath is modeled as $N_s$ spherical active Brownian particles with radius $r_s = 4$ nm. We employ a ``sticker-spacer'' framework ~\cite{zhang2024exchange}, where monomers serve as stickers of type A and type B, linked by finitely extensible springs that act as implicit spacers~\cite{kremer1990dynamics} (Supp. Fig.~\ref{fig:model_sim}). The implicit spacers are modeled using the following spring potential:
\begin{equation}
V_{\textrm{b}}(r_{ij}) = - \frac{1}{2} k_{b} R_0^2 ln \left[ 1 - \left( \frac{r_{ij}}{R_0} \right)^2\right], \hspace{5mm}  r_{ij} < R_0
\label{eq:bond}
\end{equation}
where $r_{ij}$ is the distance between the $i^{th}$ and $j^{th}$ stickers on the polymer chain, $k_b$ is the spring constant, and $R_0^2$ is the maximum extension of the bond. The interactions between monomers are determined by their types. Monomers of different type (A–B pairs) interact via a soft attractive potential that models specific, one-to-one binding with affinity strength, $\epsilon_{AB}$. The soft attractive potential is defined as:
\begin{equation}
V_{\text{A}}(r_{ij}) = - \epsilon_{AB}  \left( 1 + cos \frac{\pi r_{ij}}{r_c}\right),  \hspace{5mm}  r_{ij} < r_c
\label{eq:soft}
\end{equation}
where $\epsilon_{AB}$ is the strength of the attractive interaction or the binding affinity and $r_c$ is the cutoff distance. In contrast, monomers of the same type (A–A and B–B) interact through a purely repulsive Weeks--Chandler--Andersen (WCA) potential~\cite{weeks1971role}, which prevents nonspecific aggregation and polymer collapse. The WCA potential is given by:
\begin{equation}
V_{\text{R}}(r_{ij}) =
\begin{cases}
4\epsilon_{AA} \left[ \left( \dfrac{\sigma_{ij}}{r_{ij}} \right)^{12} - \left( \dfrac{\sigma_{ij}}{r_{ij}} \right)^6 \right] + \epsilon_{AA}, & r_{ij} \leq 2^{1/6}\sigma_{ij} \\
0, & r_{ij} > 2^{1/6}\sigma_{ij}
\end{cases}
\label{eq:wca}
\end{equation}
where $r_{ij}$ is the distance between the $i^{\text{th}}$ and $j^{\text{th}}$ particles (polymers or the bath particles), $\sigma_{ij} = \frac{\sigma_i + \sigma_j}{2}$ is the interaction range with $\sigma_{i(j)}$ denoting the diameter of the $i^{\text{th}}$ ($j^{\text{th}}$) particle, and $\epsilon_{AA}$ is the interaction strength. \\

\noindent We consider $k_b = 0.15 \, k_B T/\mathrm{nm}^2$, $R_0 = 10 \, \mathrm{nm}$, $\epsilon_{AB} = 4\, k_B T$ (or $5\, k_B T$), and $r_c = 1\, \mathrm{nm}$, with particle diameters $\sigma = 2\, \mathrm{nm}$ for polymers and $8\, \mathrm{nm}$ for bath particles in all simulations (Table~\ref{tab:parameters_sim}). The system contains $N = 1000$ polymers in a simulation box of dimensions $800 \times 100 \times 100 \, \mathrm{nm}^3$ with periodic boundary conditions. This corresponds to a dilute regime with a volume fraction of 0.0465, and number densities of $15 \times 10^{-4}$ and $15 \times 10^{-5}$ for polymers and bath particles, respectively. The dynamics evolve according to the overdamped Langevin equation:
\begin{equation}
\gamma \, \dot{\bm{r}}_i = - \nabla_i V(\bm{r}_{1},\bm{r}_{2},..., \bm{r}_{N}) + \bm{f}_{i}(t) + \bm{F}_{a} \label{eq:langevineq}
\end{equation}
where $\bm{r}_i$ is the position of the $i$th particle, $\gamma$ is the friction coefficient, and $V(\bm{r}_1, \ldots, \bm{r}_N)$ includes all bonded and non-bonded interactions. The term $\bm{f}_{i}(t)$ is a Gaussian thermal noise satisfying the fluctuation–dissipation theorem, and $\bm{F}_{a}$ represents active forces from the bath. Simulations are conducted at $T = 300\, \mathrm{K}$, with a damping timescale $m/\gamma = 10\, \mathrm{ns}$ and integration timestep, $dt = 5 \times 10^{-4}\, \mathrm{ns}$. \\ 

\begin{table}[h]
\centering
\caption{Parameters used in the simulations.}
\label{tab:parameters_sim}
  \begin{tabular*}{0.48\textwidth}{@{\extracolsep{\fill}}lll}
    \hline
    Parameters & Numerical Value \\
    \hline
    Number of polymer monomers ($N_m$) & 12  \\
    Number of polymers ($N$) & 1000  \\
    Number of bath particles ($N_s$) & 1200  \\
    Radius of a polymer monomer ($r_p$) & 1 nm  \\
    Radius of a bath particle ($r_s$) & 4 nm  \\
    Simulation box dimension ($V_\mathrm{box}$) & 800 $\times$ 100 $\times$ 100 $\mathrm{nm}^3$\\
    Total volume fraction ($\phi_{pack}$) & 0.0465 \\
    Pecl\`et  number ($Pe$) & 0, 0.5, 1.0, 2.0 \\
    Temperature ($T$) & 300 K \\
    Spring constant ($k_b$) & 0.15 $k_BT/\mathrm{nm}^2$ \\
    Maximum extent of the bond ($R_0$ ) & 10 nm \\
    Binding affinity ($\epsilon_{AB}$) & 4, 5 $k_BT$ \\
    Cutoff for soft attraction  ($r_c$) & 1 nm \\
    WCA interaction strength ($\epsilon_{AA}$) & 1 $k_BT$ \\ 
    Damping timescale ($\frac{m}{\gamma}$) & 10 ns \\
    Total simulation time ($\tau_{\mathrm{sim}}$) & 40000 $\tau$ \\
    Simulation timestep ($d\tau$) & $ 5 \times 10^{-4}$ ns \\
    \hline
  \end{tabular*}
\end{table}

\noindent We initialize the simulation by packing polymers and bath particles into the box, followed by a gradual increase in the attractive interaction between A and B stickers from 0 to $\epsilon_{AB}$ over $25 \times 10^6$ integration steps. The system is then relaxed for an additional $50 \times 10^6$ steps to allow separation into dense and dilute phases. Particle positions are recorded every $5 \times 10^4$ steps. We perform five independent simulation replicates for each combination of activity and binding affinity parameters. All reported quantities are averaged over these replicates. \\

\begin{figure*}[h!]
    \centering
    \includegraphics[width=0.95\textwidth]{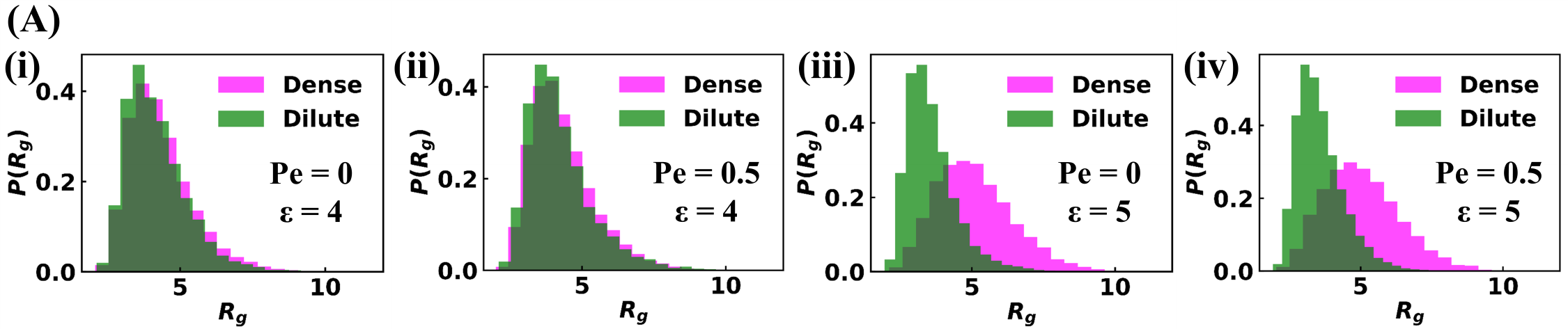}
    \caption{\textbf{Polymer compactness distinguishing dilute and droplet-like states.} Distribution of the polymer radius of gyration ($R_g$) for $Pe = 0$ and $Pe = 0.5$ at two different binding affinities: (i, ii) $\epsilon_{AB} = 4$, where clustering is absent, and (iii, iv) $\epsilon_{AB} = 5$, where clustering is present.}
    \label{fig:Rg_sim}
\end{figure*}

\noindent To further assess how the formation of self-assembled structures influences polymer conformation, we analyze the distribution of the radius of gyration ($R_g$) in both dense and dilute phases (Fig.~\ref{fig:Rg_sim}). For a polymer chain, the radius of gyration is defined as $R_g = \sqrt{\frac{1}{N_m}\sum_{i=1}^{N_m} \left( \bm{r}_i - \bm{r}_{\text{com}} \right)^2 }$, where $N_m = 12$ is the total number of monomers in the chain, $r_i$ is the position of $i^{th}$ monomer, and $r_\text{com}$ is the center of mass of the polymer chain. We consider two regimes of binding affinity: one where clustering is weak and another where robust droplet formation occurs. At low binding affinity ($\epsilon_{AB} = 4$), $R_g$ distributions are nearly identical across phases, regardless of activity, consistent with the lack of well-defined assemblies and the limited incorporation of bath particles into these phases (Fig.~\ref{fig:Rg_sim}(i, ii)). This implies minimal preferential interactions between polymers and bath particles, suggesting that polymer-polymer interactions dominate over polymer-bath interactions within these weak clustered regions. In contrast, at higher binding strength ($\epsilon_{AB} = 5$), prominent self-assembled structures emerge, and polymers in the dense phase adopt more extended conformations than those in the dilute phase (Fig.~\ref{fig:Rg_sim}(iii, iv)). This is reflected in broader $R_g$ distributions and a shift of the distribution peak toward higher $R_g$ values in the dense phase. In this regime, the dense phase consists of large clusters that efficiently incorporate bath particles. To accommodate the larger size of bath particles, polymers adopt more extended conformations within the dense phase. These findings highlight that activity, through enhanced steric interactions with bath particles, facilitates polymer expansion and promotes the growth of dense clusters.

\end{document}